\documentstyle [12pt] {article} \textwidth 160mm\textheight
220mm\topmargin - 5mm \oddsidemargin -5mm\evensidemargin -5mm
\begin {document} %\footskip=1000mm

\begin {center} {\large \bf Comment to the paper " The energy
conservation law\\ for electromagnetic field in application to problems
\\of radiation of moving particles "} \end {center}

\vskip 1mm \begin {center} {\large E.G.Bessonov} \end {center}

In the paper \cite {bolot} the energy conservation law (the Poynting
theorem) was applied to a problem of  radiation of a charged particle
in an external electromagnetic field. The authors consecutively and
mathematically strictly solved the problem but received wrong result.
They derived an expression which includes a change of the energies of
the electromagnetic fields accompanying the homogeneously moving
particle $\Delta W = W_2 - W_1$ corresponding to the initial and final
velocity of the particle (see expression (19) in \cite {bolot}).  The
energy of the field accompanying the particle $W$ is the energy of the
particle of the electromagnetic origin. It should not enter the
solution of the problem. The authors do not specify the dimensions of
the particle. For pointlike particle this energy and the change $\Delta
W$ are infinite values and consequently the expression (19) loses
sense. In quantum theory the derived expression require some
renormalization. In classical theory in the section devoted to the
energy conservation law the energy of the accompanying field that is
the energy of the particle of the electromagnetic origin is hidden in
the total energy of the electromagnetic origin and hence it is appeared
unnoticed. For this reason the solutions based on the use of the energy
conservation law lead to the wrong results when $\Delta W \ne 0$ \cite
{bes1}. The received solutions differ from the solutions based on the
equations of motion of particles in the external fields.

We will explain our observation using the second example considered by
the authors. This example was formulated as follows. Let a charged
particle is moving in a positive direction of the axis "z" with a
velocity $\vec v_1$. In some area with the linear dimensions $L$
located near to the origin of the reference frame the external
electromagnetic fields are created where the velocity of the particle
is varied under some law which is not specified.  Then the particle go
out from this area and it's velocity accepts the value $\vec v_2$ which
hereinafter is not changed. The authors proceed from the expression for
the energy conservation law of the form

            \begin {equation} %1
            {\partial\over \partial t} \int _V {| \vec E | ^2 + | \vec
            H | ^2\over 8\pi} d V = - \int _V \vec \jmath \vec EdV -
      {C\over 4\pi} \int _S [\vec E \vec H] d\vec S, \end {equation}
where $\vec E$, $\vec H$ are vectors of the electric and magnetic
fields respectively created in a general case by a set of particles,
charged bodies and magnets, $\vec \jmath$ a vector of density of a
current, the sign $V$ under the integral means that the integral is
carried out through a chosen volume $V$ and the sign $S$ means that the
integral is carried out through a surface $S$ limiting this volume.
This law (the Poynting theorem) follows from the Maxwell's equations.

From this law the authors came to the expression

            \begin {equation} %2
            {c\over 4\pi} \int _ {-\infty} ^ {+ \infty} dt \int _S
            [\vec E ^ {"} \vec H ^ {"}] d\vec S = - \int _ {-\infty}
            ^ {+ \infty} dt \int _V \vec \jmath \vec EdV -
            \Delta W, \end {equation}
where the vectors $\vec E^ {"} $, $\vec H^ {"} $ are vectors of free
electric and magnetic fields emitted by a particle, $W_1 = (1/8\pi)
\int (| \vec E_1 | ^2 + | \vec H_1 | ^2) d V $, $W_2 = (1/8\pi) \int
( | \vec E_2 | ^2 + | \vec H_2 | ^2) d V $ are the total energies of the
electromagnetic fields, created by the homogeneously moving charged
particle in the unlimited space (the energies of the accompanying
field), vectors $\vec E_1$, $\vec H_1$ and $\vec E_2$, $\vec H_2$ are
the vectors of the electric and magnetic field strengths created by a
particle moving homogeneously with velocities $\vec v_1$, $\vec v_2$
accordingly. It is supposed that the boundary of the volume $V$ is
chosen so far that the wavepacket of radiation was in time to be
separated from the field of the charged particle so that free fields of
radiation and the field accompanying the particle are not overlapped.

Further the authors go to the conclusion that the flow of radiation
from the volume $V$ according to (2) is determined not only by the
work of forces acting on the charged particles by fields (integral
from $\vec \jmath \vec E$) but also by change of the energy of the
accompanying electromagnetic field $\Delta W$.

Now we notice that the vector of the electric field strength in the
region of location of the particle can be presented in the form $\vec E
= \vec E_ {ext} + \vec E _ {s} $, where $\vec E_ {ext} $ is the vector
of the external electric field strength created by charged bodies and
other particles, $\vec E _ {s} $ vector of the electric field
strength produced by the particle under consideration. Therefore the
external fields and the fields produced by a particle (inertial and
radiating self-fields) were took into account in the change of the
energy of the particle $\varepsilon$ and the value $\int _V \vec \jmath
\vec E d V = d\varepsilon /dt$ \cite {landau}, \cite {jackson}. That is
why the value $\int _ {-\infty} ^ {+ \infty} dt \int _V \vec \jmath
\vec EdV = \Delta \varepsilon = \varepsilon _2 - \varepsilon _1 = mc^2
(\gamma _2 - \gamma _1) $ in the equation (2) is the change of the
total energy of the particle where $m$ is the weight of the particle,
$\gamma = 1/\sqrt {1 - \beta ^2} $, $\beta = | \vec v/c | $, the
subscripts $1,2$ are related to initial and final velocity of the
particle. The value ${c/4\pi} \int _ {-\infty} ^ {+ \infty} dt \int _S
[\vec E ^ {"} \vec H ^ {"}] d\vec S = \varepsilon ^ {rad} $ is the
energy of the electromagnetic radiation emitted by the particle in the
form of free electromagnetic waves.  Thus the expression (2) can be
presented in the form

            \begin {equation} %3
            \Delta \varepsilon = - \varepsilon ^ {rad} - \Delta W.
            \end {equation}

In the presented example it was supposed that the external fields are
static and the energy of these fields is constant. In a static case the
electric field is potential one $\int \vec E_ {ext} (\vec r) d \vec r
= 0$. The external field could be the magnetic one. Therefore
obviously the change of the energy of the particle in the case of
static fields should be equal to the energy of radiation taken with the
negative sign

            \begin {equation} %4
            \Delta \varepsilon = - \varepsilon ^ {rad}.
            \end {equation}

The expression (4) follows also from the equations of motion of the
particle in the external fields taking into account the radiation
reaction force and the laws of radiation of a particle in the external
fields which are determine the rate of losses of the energy of the
particle in the form of radiation.

Contrary to expected result a superfluous term has appeared in the
expression (3)  which is equal to the change of the energy of the field
$\Delta W$ accompanying the particle and differ from zero as the initial
$v_1 = | \vec v _1 | $ and final $v_2 = | \vec v _2 | $ velocities are
not equal ($v_1 - v_2 \ne 0$).

The presence of the superfluous term $\Delta W$ is in accordance with
the conclusions made in the paper \cite {bes1} that from the equations
of Maxwell-Lorentz does not follow the correct energy conservation law
that is the law which describe the nature correct way since the
equations of Maxwell and equations of Lorentz are inconsistent. The
expression (1) contains a logic error consisting in the fact that in
the first field term of this expression the energy of the
electromagnetic field is included and in this energy the energy of the
accepted electromagnetic field of the particle i.e. the energy of the
particle of electromagnetic origins is hidden. It means that the energy
of the particle of the electromagnetic origin in the equation (1) is
presented in two terms (left and first right term). Accordingly the
energy of the particle of the electromagnetic origin in the equation
(3) is also presented in two terms ($\Delta \varepsilon$ and $\Delta
W$). We should like to remind that energy of the particle is a sum
of energies of the electromagnetic and non-electromagnetic origin and
in the case of pointlike particles they are infinite and have opposite
sign and their sum presents the experimentally observable value
$\varepsilon $ \cite {landau}. Thus the energy of particles of the
electromagnetic origin is presented in expressions (1), (3) twice and
that is why the Poynting theorem generalized on a case of a system of
fields and particles becomes incorrect. In the case of pointlike
particles the value $\Delta W$ in the expression (3) is infinite when
the value $v_1 - v_2 \ne 0$ and that is why this expression loses its
sense. The logic error consist in the double inclusion of the energy
of the particle of the electromagnetic origin in the same equation.

We would like to remind the energy conservation law for a system of the
electromagnetic field and particles in an integral form $\partial
\varepsilon ^ {\Sigma} / \partial t = 0$ or $\varepsilon ^ {\Sigma} =
const$ where

            \begin {equation} %5
            \varepsilon ^ {\Sigma} = \int {| \vec E | ^2 + | \vec
            H | ^2\over 8\pi} d V + \Sigma \varepsilon _i,
            \end {equation}
$\varepsilon _i$ is the energy of a particle $i$, and the integration
is carried out through the whole space \cite {landau}. The first term
from the right in the expression (5) contains both the free field of
radiation emitted by charged particles and the field accompanying these
particles. The dimensions, charge, and weight of the particle and also
their structure can be arbitrary. That is why massive charged bodies and
magnets can enter in (5). Massive bodies can have complex structure. In
this case the exchange of the energy of electromagnetic fields is
possible in internal degrees of freedom of a body (for example, in
heating the body). At that the weight and, accordingly, energy of
bodies $\varepsilon _i$ will be increased.

In a general case if the wave packet of radiation emitted by a system
of particles will be in time to be separated from the fields
accompanying these particles then the change of the energy of the
system of particles $\Delta \varepsilon = \Sigma \Delta \varepsilon _i$
and the change of the energy of the electromagnetic field according to
(5) will be determined by the same expression (3) where now
$\Delta W$ is the change of the energy of the accompanying
electromagnetic fields of all particles\footnote {Certainly it is
possible to receive this conclusion proceeding from the expression
(1).}. It means that the conclusion about a logic mistake made at the
proof of the energy conservation law for a system of electromagnetic
field and particles made in the paper \cite {bes1} for a general case
non-obviously was confirmed by the authors of the commented paper in
their example.

At the derivation of the energy conservation law for the system of the
electromagnetic field and particles a mistake was made which further
was accepted by repetition in many papers and textbooks. Therefore the
interpretation of this law in the textbooks should be changed and
should be treated in the form of an open question in classical
electrodynamics.

The author thanks B.M.Bolotovskii and S.N.Stoliarov for useful
discussions of the present comment.

\end {document}